# SPONTANEOUS EMISSION OF A QUANTUM PARTICLE UNDER STRONG STARK INTERACTION WITH RESONANT VACUUM FIELD

A.M.Basharov

*RRC «Kurchatov Institute»*, e-mail: basharov@gmail.com

It has been shown that strong Stark interaction of a quantum particle with a vacuum electromagnetic field reduces the speed of the one-quantum spontaneous radiation and leads to additional shift of frequency of radiation transition.

## 1. Introduction

The ordinary atom in a vacuum electromagnetic field with zero photon density undergoes the Lamb shift of levels and spontaneous transition from the excited to the ground state [1]. The description of spontaneous decay of the excited state is successfully given by the two-level model of a quantum particle [2]. In conventional conditions the Lamb shift of resonant frequency of atom in a vacuum electromagnetic field is negligibly small in comparison with the characteristic Rabi frequency describing radiation transitions between quantum levels. Therefore, in the dynamics of spontaneous decay the Lamb shift is taken into account only in the renormalization of resonant transition frequency and does not influence the decay in any way. In particular, the speed of spontaneous decay of atom does not depend on it [1,2]. The Lamb shift infinitesimality involved is due to the fact that the Rabi frequency is of the first order value of the coupling constant with an electromagnetic field whereas the Lamb shift is of the second order of this constant. Meanwhile, the two-level model of atom successfully describes not only the behavior of atom in the conventional medium, but also in artificial media such as a photon crystal [3], and also describes dynamics of various artificial emitters similar to quantum dot [4] or atom-photon cluster [5]. A distinctive feature of cases described in works [3-5] is that they represent a variety of two-quantum processes; with the characteristic Rabi frequencies becoming of the second order of the coupling constants of a quantum particle with either two electromagnetic fields or one electromagnetic field and medium particles. The second order of the coupling constant Rabi frequencies becomes very small. Thus, there are cases in which the values of the second order of the coupling constants (such as Lamb frequency shift) may prove to be of the same order as that of the Rabi frequency describing spontaneous decay. Moreover, for multiphoton decays the values of the second order of the coupling constants will exceed the characteristic Rabi frequency. Therefore, the study of spontaneous emission at low Rabi frequency is of interest. This paper has provided the solution to the problem of spontaneous emission of a two-level quantum particle in a photon-free vacuum electromagnetic field under the conditions when the processes described by the second order of the coupling constant with electromagnetic fields are truly essential. Thus we consider that the direct transition between excited and ground states of the particle is performed by emitting one photon. In the study the so-called Stark interaction of a two-level particle with a quantized electromagnetic field has been taken into account for the first time. In a classical field the Stark interaction is known as a high-frequency Stark effect [6], which also generates level shifts of a quantum particle, with its value being proportional to the intensity of the classical field [6]. In a quantized field the classical Stark level shift is replaced by the operator called the Stark interaction operator. It is of the same second order of the coupling constant with quantized electromagnetic field, as that of the Lamb shift of levels. The interpretation of the Lamb shift as part of the total Stark shift was discussed in ref.[1]. The part of the Stark interaction operator remaining after the separation of the Lamb shift is always neglected in previous works because its average value is equal to zero in the case of zero photon density of a vacuum electromagnetic field. That is why the Stark interaction was not taken into account earlier.

In the present paper it is established that in the Markov approximation the part of the Stark interaction with zero average (shortly speaking the Stark interaction) makes an essential impact on dynamics of a two-level particle. In particular, it suppresses spontaneous emission and leads to additional shift of frequency of resonant transition. Our model is applicable to various two-quantum transitions in radiation processes described in works [3-5]. As the general model of spontaneous radiation of a quantum particle has been used in our study, the results obtained can affect a variety of various optical effects and methods, for example in spectroscopy, in superradiation, etc.

The quantum stochastic differential equations [7] used widely in quantum optics [8,9] represent the investigation method of spontaneous emission under the strong Stark interaction with a resonant vacuum field. From the point of view of the quantum stochastic differential equations the direct one-photon transitions between quantum particle levels are described by quantum Wiener process, which allowed deriving kinetic equations and relaxation operator for a two-level atom [8,9] in a straightforward and elegant way by neglecting the Stark interaction operator. In works [8,9] quantum stochastic equations were of the so-called Langevin type. Our approach in using the quantum stochastic differential equations is different to refs.[8,9] through involving the gauge (or number) process [7], which has been applied for describing the Stark interaction. As a result, the quantum stochastic equations appeared to be of non-Langevin type. The investigation surveys how to obtain such non-Langevin quantum stochastic equations and how to simply deduce the kinetic equations for a quantum particle. Our consideration of the Stark level shifts can appear useful for solving other similar problems in which it is necessary to take into account the operator represented by the gauge (or number) process.

## 2. Formulation of the problem

The effective Hamiltonian $H^{Eff}$ of the system consisting of a two-level quantum particle and resonant quantized electromagnetic field represents the sum of the Hamiltonian of the "isolated" two-level quantum particle $H^{TL}$, the Hamiltonian of an electromagnetic field $H^F$, the interaction operator with the electromagnetic field describing the direct transitions between particle levels with the emission or absorption of a photon $H^{Tr}$, and of the Stark interaction operator describing the virtual transitions with returning to initial level without photon emission or absorption $H^{St}$ [5]:

$$H^{Eff} = H^{TL} + H^F + H^{Tr} + H^{St}, \qquad (1)$$

$$H^{TL} = \hbar\omega_{21} R_3, \quad H^F = \int \hbar\omega b_\omega^+ b_\omega d\omega,$$

$$H^{Tr} = \int \Gamma_\omega b_\omega^+ D_{12}(\omega) R_- d\omega + \int \Gamma_\omega b_\omega D_{21}(\omega) R_+ d\omega, \quad H^{St} = \int \Gamma_\omega \Gamma_{\omega'} b_\omega^+ b_{\omega'} \Pi(\omega,\omega') R_3 d\omega d\omega'.$$

Here, the operators $R_3 = (|E_2><E_2| - |E_1><E_1|)/2$, $R_+ = |E_2><E_1|$, $R_- = |E_1><E_2|$ realize two-dimensional irreducible representation of $su(2)$-algebra with the commutation relations $[R_3, R_\pm] = \pm R_\pm$, $[R_+, R_-] = 2R_3$; $|E_1>$ and $|E_2>$ are the ground and excited quantum states of the two-level particle, $\omega_{21}$ is the transition frequency between them with the account of the Lamb shift. The quantized electromagnetic field is characterized by the central frequency $\Omega_\Gamma$, annihilation and creation operators ($b_\omega$ and $b_\omega^+$) of photons with the commutation relations $[b_\omega, b_{\omega'}^+] = \delta_{\omega\omega'}$. The coupling constant of a quantum particle with the electromagnetic field is designated as $\Gamma_\omega$. Parameters $D_{12}(\omega)$ and $\Pi(\omega,\omega')$ characterize the Rabi frequency and the Stark shift of levels. They depend on the very emitter whose dynamics is modeled by a two-level quantum particle.

For the one-quantum resonance $\Omega_\Gamma \approx \omega_{21}$ of the quantized electromagnetic field with optically allowed transition $|E_2> \to |E_1>$ in an ordinary atom, $D_{21}(\omega) = d_{21}$, where $d_{21}$ is the matrix element of the atomic dipole moment operator. The value $\Pi(\omega,\omega')$ is expressed with the parameters of the multi-photon resonance theory [5,6],

$$\Pi(\omega,\omega') = \tfrac{1}{2}\{\Pi_2(\omega) + \Pi_2(\omega') - \Pi_1(\omega) - \Pi_1(\omega')\}.$$

The equation below gives the notion of their value [5,6]

$$\Pi_k(\nu) = \sum_{i,jk} \frac{|d_{kj}|^2}{\hbar}\left(\frac{1}{\omega_{kj}+\nu} + \frac{1}{\omega_{kj}-\nu}\right),$$

where $d_{kj}$ is the matrix element of the dipole moment operator for the transition $|E_k> \to |E_j>$ of the frequency $\omega_{kj}$ in the atom under consideration.

The role of the Stark effect for an ordinary atom will only be displayed at a very low value $d_{12}$. At $\Pi(\omega,\omega') = 0$ and $D_{12}(\omega) = d_{12}$ the effective Hamiltonian (1) describes the conventional dynamics of a two-level atom, which had been widely considered in different textbooks and monographs, including [2,6,9].

For the two-quantum resonance the Stark effect will always be essential, e.g. in the case of the Raman resonance $\Omega_\Gamma - \omega_c \approx \omega_{21}$ of a broadband quantized electromagnetic field with the central frequency $\Omega_\Gamma$, cavity mode of frequency $\omega_c$ and spectral width $\Delta\omega_c$ with the optically forbidden transition $|E_2> \to |E_1>$ of an atom in microresonator. In this case one can introduce the notion of the atom-photon cluster [5]. Then, $H^{St}$ will look the same as that of an ordinary atom, and $D_{21}(\omega) = -g\Pi_{21}(\omega)$, where $g \sim \Gamma_{\omega_c}\Delta\omega_c$ is the coupling constant of atom with the cavity mode, and $\Pi_{21}(\omega) = \sum_j \frac{d_{2j}d_{j1}}{\hbar}\left(\frac{1}{\omega_{j2}+\omega} + \frac{1}{\omega_{jm}-\omega}\right)$ is the parameter of the two-quantum resonance describing the related transition [5,6]. For the atom-photon cluster, the operators $R_3$ and $R_\pm$ are the generators of the third order polynomial algebra, the two-dimensional representation of which is reduced to the generators of the su(2)-algebra for the simplest case of single atom-photon cluster excitation [5].

It is to be emphasized again that the influence of the Stark effect for the atom-photon cluster remains essentially important for the spontaneous emission as the operators $H^{Tr}$ and $H^{St}$ are of one and the same second order of the coupling constant with the electromagnetic field.

The thermal motion of atoms, recoil effects in the processes of emission/absorption of photons, level degeneracy and polarization states of photons are neglected in the Hamiltonian (1). It is assumed that the quantum particle is located in the spatial region near the point $\bar{r}=0$. Besides, for simplicity of the effective Hamiltonian representation in terms of su(2)-generators, the term given below is neglected

$$\sum_\omega \Gamma_\omega b_\omega^+ \sum_{\omega'} \Gamma_{\omega'} b_{\omega'} \{\frac{\Pi_1(\omega)+\Pi_1(\omega')+\Pi_2(\omega)+\Pi_2(\omega')}{4}\{|E_2><E_2| + |E_1><E_1|\}.$$

It does not affect the spontaneous emission of a quantum particle within the framework of the Stark effect, which is part and parcel of the current investigation.

The system comprising the quantum particle and the broadband electromagnetic field is given by the Schrödinger equation for the state vector $|\Psi>$

$$i\hbar \frac{d}{dt}|\Psi> = H^{Eff}|\Psi> \tag{2}$$

with the initial vector $|\Psi_0>$ which is the factorization of the quantum particle state $|\Psi_0^{TL}>$ and electromagnetic field state $|\Psi_0^F>$

$$|\Psi_0> = |\Psi_0^{TL}> \otimes |\Psi_0^F>. \tag{3}$$

The electromagnetic field is the thermostat with the central frequency $\Omega_\Gamma$ and zero photon density

$$<\Psi_0^F | b_\omega b_{\omega'}^+ | \Psi_0^F> = \delta(\omega-\omega'),\tag{4}$$

$$<\Psi_0^F | b_\omega | \Psi_0^F> = <\Psi_0^F | b_\omega^+ | \Psi_0^F> = <\Psi_0^F | b_\omega^+ b_{\omega'} | \Psi_0^F> = <\Psi_0^F | b_\omega b_{\omega'} | \Psi_0^F> = <\Psi_0^F | b_\omega^+ b_{\omega'}^+ | \Psi_0^F> = 0.$$

## 3. Schrödinger equation for the evolution operator as a non-Langevin quantum stochastic equation

The Schrödinger equation (2) can be conveniently rewritten in the interaction representation as the equation for the system evolution operator $U$ in the form

$$i\hbar \frac{d}{dt}U(t) = (H^{Tr}(t) + H^{St}(t))U(t).\tag{5}$$

$$H^{Tr}(t) = \int \Gamma_\omega b_\omega^+ e^{i(\omega-\omega_{21})t} D_{12}(\omega) R_- d\omega + \int \Gamma_\omega b_\omega e^{-i(\omega-\omega_{21})t} D_{21}(\omega) R_+ d\omega,\tag{6}$$

$$H^{St} = \int \Gamma_\omega \Gamma_{\omega'} b_\omega^+ e^{i(\omega-\omega_{21})t} b_{\omega'} e^{-i(\omega'-\omega_{21})t} \Pi(\omega,\omega') R_3 d\omega d\omega'\tag{7}$$

with the formal solution in terms of the $\tilde{T}$ - exponent

$$U(t) = \tilde{T}\exp\left(\frac{1}{i\hbar}\int_0^t (H^{Tr}(t') + H^{St}(t'))dt'\right).\tag{8}$$

Assume that the coupling parameter $\Gamma_\omega$ along with the parameters of the direct transition and the Stark level shifts are not affected by the frequency $\omega$

$$\Gamma_\omega = const = \Gamma_{\omega_{21}},\ \Pi(\omega,\omega') = const = \Pi(\omega_{21},\omega_{21}),\ D_{12}(\omega) = const = D_{12}(\omega_{21}).\tag{9}$$

The following operators can be introduced

$$b(t) = \frac{1}{\sqrt{2\pi}}\int_{-\infty}^{\infty} d\omega e^{-i(\omega-\omega_{21})t} b_\omega,\ b^+(t) = \frac{1}{\sqrt{2\pi}}\int_{-\infty}^{\infty} d\omega e^{i(\omega-\omega_{21})t} b_\omega^+,$$

$$B(t) = \int_0^t dt' b(t'),\ B^+(t) = \int_0^t dt' b^+(t'),\ \Lambda(t) = \int_0^t dt' b^+(t')b(t').\tag{10}$$

Let's present the equation for the evolution operator in a dimensionless form with dimensionless time $\tau$ and parameters $\chi$ and $\eta$ describing direct quantum transitions between particle levels and the Stark interaction,

$$idU(t) = \{\chi R_+ dB(\tau) + \chi R_- dB^+(\tau) + \eta R_3 d\Lambda(\tau)\}U(\tau),\tag{11}$$

$$dB(\tau) = B(\tau+d\tau) - B(\tau),\ dB^+(\tau) = B^+(\tau+d\tau) - B^+(\tau),\ d\Lambda(\tau) = \Lambda(\tau+d\tau) - \Lambda(\tau),\tag{12}$$

supposing that the integration limits in (6) and (7) range from $-\infty$ to $+\infty$, rather than from 0 to $+\infty$. The difference in the notation for dimensionless and dimensional increments of the basic stochastic processes is designated by their time argument.

The specified expansion of integration limits in (6) and (7) as well as conditions (9) define the interaction of a quantum particle with the quantized electromagnetic field (4) as the Markov process [7-9]. Thus, differentials (12) are the increments of the quantum Wiener and gauge processes [7] satisfying the Hudson-Parthasarathy algebra [7]:

$$d\Lambda(\tau)d\Lambda(\tau) = d\Lambda(\tau),\ d\Lambda(\tau)dB^+(\tau) = dB^+(\tau),\ dB(\tau)d\Lambda(\tau) = dB(\tau),\ dB(\tau)dB^+(\tau) = d\tau.\tag{13}$$

$$d\Lambda(\tau)dB(\tau) = d\Lambda(\tau)d\tau = dB^+(\tau)d\Lambda(\tau) = dB^+(\tau)d\tau = dB(\tau)d\tau = 0.$$

For the Markov process the integrals in equation (8), as well as equation (5) become mathematically undefined. Integrals in (8) are interpreted to mean in the Ito sense [6-9]. Obeying the analogous procedure described in work [10], it is not difficult to derive the following quantum stochastic Ito equation from eq.(8) for the increment $dU(\tau) = U(\tau + d\tau) - U(\tau)$ of the evolution operator

$$dU(t) = \{\exp(-i(\chi R_+ dB(t) + \chi R_- dB^+(t) + \eta R_3)d\Lambda(t))) - 1\}U(t).$$

Expansion of the exponent with account of the Hudson-Parthasarathy algebra (13) gives

$$dU(\tau) = \chi^2 R_+ R_- \frac{e^{i\eta} - 1 - i\eta}{\eta^2} d\tau U(t) - \chi R_+ \frac{e^{i\eta} - 1}{\eta} dB(\tau)U(\tau) - \chi R_- \frac{e^{i\eta} - 1}{\eta} dB^+(\tau)U(\tau) + \quad (14)$$

$$+ \{\cos(\eta) - 1 - i\sin(\eta)R_3\}d\Lambda(\tau)U(\tau).$$

In the absence of the Stark interaction $\eta = 0$, equation (14) coincides with the case studied in refs. [6-9]. The dependence of the evolution operator on the gauge process $d\Lambda(\tau)$ is the characteristic feature of the non-Langevin process [10].

## 4. Suppression of spontaneous decay by the Stark Interaction

The equation for the system density matrix $\rho(\tau) = U(\tau) | \Psi_0 \rangle \langle \Psi_0 | U^+(\tau)$ is generated by the Ito equation (14) for the evolution operator by calculating the increment, which is the typical manifestation of the stochastic differential equation techniques

$$d\rho(\tau) = \rho(\tau + d\tau) - \rho(\tau) =$$
$$= dU(\tau) | \Psi_0 \rangle \langle \Psi_0 | U^+(\tau) + U(\tau) | \Psi_0 \rangle \langle \Psi_0 | dU^+(\tau) + dU(\tau) | \Psi_0 \rangle \langle \Psi_0 | dU^+(\tau).$$

Derived from eq.(14) and the Hudson-Parthasarathy algebra (13), the expression for the density matrix $\rho(\tau)$ of the entire system is too bulky to be treated in this work. However, it follows from it that the equation for the density matrix of the two-level quantum particle $\rho^{TL}(\tau) = Tr_F \rho(\tau)$ should be taken in the form:

$$\frac{d\rho^{TL}}{d\tau} = -\chi^2 R_+ R_- \left(\frac{1-\cos\eta}{\eta^2} + i\frac{\eta - \sin\eta}{\eta^2}\right)\rho^{TL} - \chi^2 \rho^{TL} R_+ R_- \left(\frac{1-\cos\eta}{\eta^2} - i\frac{\eta - \sin\eta}{\eta^2}\right) + \quad (15)$$

$$+ 2\chi^2 \frac{1-\cos\eta}{\eta^2} R_- \rho^{TL} R_+.$$

Here, $R_+ R_- = | E_2 \rangle \langle E_2 |$, the operator $\chi^2 \frac{\sin\eta - \eta}{\eta^2} R_+ R_-$ describes an additional shift of the decaying level due to the Stark interaction $H^{St}$ of the quantum particle with the vacuum electromagnetic field, and $L = \chi \frac{\sqrt{1-\cos\eta}}{\eta} R_-$ is the Lindblad generator.

The additional shift $\chi^2 \frac{\sin\eta - \eta}{\eta^2}$ of the decaying level $| E_2 \rangle$ differs from the Lamb shift and is due to the very process of its decay. This shift is also different from the conventional notion of the Stark shift $\langle \Psi_0^F | H^{St} | \Psi_0^F \rangle$, as the latter one is equal to zero in the case of zero photon density (4). The shift of the decaying level is referred to as the decaying Stark shift. Emphasize that in the absence of radiation transition with the emitting a photon, $\chi = 0$, the decaying Stark shift is absent. A complete picture of the non-Langevin dynamics of the two-level particle is given by the equations for matrix elements

$$\frac{d\rho_{22}^{TL}}{d\tau} = -2\chi^2 \frac{1-\cos\eta}{\eta} \rho_{22}^{TL}, \quad \frac{d\rho_{11}^{TL}}{d\tau} = 2\chi^2 \frac{1-\cos\eta}{\eta} \rho_{22}^{TL},$$

$$(16)$$

$$\frac{d\rho_{21}^{TL}}{d\tau} = -\chi^2 \left(\frac{1-\cos\eta}{\eta^2} - i\frac{\eta - \sin\eta}{\eta^2}\right)\rho_{21}^{TL}.$$

Thus, the Stark interaction $H^{St}$ gives rise to both the renormalization of the conventional radiation decay constants (with neglecting the Stark effect) and to the additional frequency shift of the resonant quantum transition. At any values of the Stark parameter $\eta$, the Stark interaction with the vacuum electromagnetic field suppresses radiation transitions in the single emitter, decreasing the constant of radiation decay. As a result, the gauge process acts as if it stabilized the excited quantum particle.

Note, that the Stark shift can be treated as a consequence of virtual transitions while absorbing and emitting the virtual photon, which leads to the returning to the same quantum level. In quantum mechanics, these transitions interfere with the real transition from the excited state to the ground state. As a result, the total transition rate to the ground state decreases. In a hypothetical situation, at a sufficiently high value of the Stark shift $\eta \to 2\pi$ a particle participates only in virtual transitions, with no transition to the ground state, that is, the Stark interaction "freezes" a particle at the excited level. Recall that the above mentioned statements are related to the case of zero photon density of the vacuum electromagnetic field.

In the absence of the direct quantum transitions between energy levels $\chi = 0$, the gauge process in the photon-free electromagnetic field does not exert influence on the state of a quantum particle.

The basic parameters $\chi$ and $\eta$ may be expressed in terms of dimensional physical values. If the frequency $\omega_{21}$ of resonant transition is taken as the characteristic frequency, and $\omega_{21}^{-1}$ as the characteristic time, then $\tau = \omega_{21} t$. For the one-quantum resonance,

$$\chi = \Gamma_{\Omega_\Gamma} \hbar^{-1} d_{12}, \quad \eta = \chi^2 \frac{\Pi_2(\omega_{21}) - \Pi_1(\omega_{21})}{2(d_{12})^2 /(\hbar \omega_{21})},$$

so that the role played by the Stark interaction will be essential under the condition

$$\frac{|\Pi_2(\omega_{21}) - \Pi_1(\omega_{21})|}{2(d_{12})^2 /(\hbar \omega_{21})} \gg 1.$$

For the two-quantum resonance of an atom in microresonator (one-quantum decay of atom-photon cluster [5]), the dimensionless parameters are expressed through dimensional values as follows

$$\chi = g \Gamma_{\Omega_\Gamma} \Pi_{21}(\Omega_\Gamma) \hbar^{-1} \sim \Gamma_{\Omega_\Gamma}^2 \Pi_{21}(\Omega_\Gamma) \hbar^{-1} \Delta \omega_c, \quad \eta = \chi^2 \hbar \frac{(\Pi_2(v) - \Pi_1(\Omega_\Gamma))}{g^2 (\Pi_{21}(\Omega_\Gamma))^2} = \chi \frac{\Pi_k(\Omega_\Gamma) \Omega_\Gamma}{\Pi_{21}(\Omega_\Gamma) \Delta \omega_c}.$$

Because $\Delta \omega_c \ll \Omega_\Gamma$, the dimensionless value $\eta$ determining the Stark interaction is of the order of unity or greater, as a rule. That is why the effect considered is of wide application in two-quantum radiation processes of quantum and nonlinear optics.